\documentclass[journal,letterpaper,twocolumn,final,10pt]{IEEEtran}

\usepackage[final]{graphicx}
\usepackage{amsmath}
\usepackage{amssymb,times}
\usepackage{cite}
\usepackage{graphicx}
\usepackage{subfigure}
\usepackage{algorithmic}
\usepackage{algorithm}
\usepackage{dsfont}
\usepackage{amsthm}
\usepackage{amsfonts}
\usepackage{mathrsfs}
\usepackage{pifont}
\usepackage{verbatim}
\usepackage{upgreek}
\usepackage{color}
\usepackage{eucal}
\usepackage{balance}

\newcommand{\bbm}{\begin{bmatrix}}
\newcommand{\ebm}{\end{bmatrix}}

\newcommand{\bit}{\begin{itemize}}
\newcommand{\eit}{\end{itemize}}

\newcommand{\ben}{\begin{enumerate}}
\newcommand{\een}{\end{enumerate}}

\newcommand{\bdesc}{\begin{description}}
\newcommand{\edesc}{\end{description}}

\newcommand{\bea}{\begin{array}}
\newcommand{\eea}{\end{array}}

\newcommand{\beqa}{\begin{eqnarray}}
\newcommand{\eeqa}{\end{eqnarray}}

\newcommand{\ds}{\displaystyle}

\newcommand{\Comment}[1]{}

\def\C{{\mathds C}}

\def\cA{\mbox{$\mathcal A$}}
\def\cB{\mbox{$\mathcal B$}}
\def\cC{\mbox{$\CMcal C$}}

\def\cN{\mbox{$\CMcal N$}}

\newcommand{\be}{\begin{equation}}
\newcommand{\ee}{\end{equation}}

\newcommand{\bzero}{{\mbox{\boldmath $0$}}}

\newcommand{\bn}{{\mbox{\boldmath $n$}}}
\newcommand{\bm}{{\mbox{\boldmath $m$}}}

\newcommand{\bor}{{\mbox{\boldmath $r$}}}

\newcommand{\bv}{{\mbox{\boldmath $v$}}}
\newcommand{\bx}{{\mbox{\boldmath $x$}}}

\newcommand{\bz}{{\mbox{\boldmath $z$}}}

\newcommand{\bI}{{\mbox{\boldmath $I$}}}

\newcommand{\bM}{{\mbox{\boldmath $M$}}}

\newcommand{\bR}{{\mbox{\boldmath $R$}}}

\newcommand{\bZ}{{\mbox{\boldmath $Z$}}}

\newcommand{\balpha}{{\mbox{\boldmath $\alpha$}}}

\newcommand{\bpi}{{\mbox{\boldmath $\pi$}}}

\newcommand{\btheta}{{\mbox{\boldmath $\theta$}}}

\newcommand{\dmax}{\begin{displaystyle}\max\end{displaystyle}}

\newcommand{\test}{\mbox{$
\begin{array}{c}
\stackrel{ \stackrel{\textstyle H_1}{\textstyle >} }{
\stackrel{\textstyle <}{\textstyle H_0} }
\end{array}
$}}

\IEEEoverridecommandlockouts

\begin{document}

\title{Novel Parameter Estimation and Radar Detection Approaches for Multiple Point-like Targets: Designs and Comparisons}

\author{Pia Addabbo, \emph{Senior Member, IEEE}, Jun Liu, \emph{Senior Member, IEEE},\\ Danilo Orlando, \emph{Senior Member, IEEE}, 
and Giuseppe Ricci

\thanks{ This work of J. Liu was supported in part by the National Key Research and Development Program of China
(No. 2018YFB1801105), the National Natural Science Foundation of China under Grant (No.
61871469), and the Youth Innovation Promotion Association CAS (CX2100060053).}
\thanks{P. Addabbo is with Universit\`a degli Studi ``Giustino Fortunato", viale Raffale Delcogliano, 12, 82100 Benevento, Italy (e-mail: p.addabbo@unifortunato.eu).}
\thanks{J. Liu is with the Department of Electronic Engineering and InformationScience, University of Science and Technology of China, Hefei 230027, China (e-mail: junliu@ustc.edu.cn). }
\thanks{D. Orlando is with Universit\`a degli Studi ``Niccol\`o Cusano'', via Don Carlo Gnocchi 3, 00166 Roma, Italy (e-mail: danilor78@gmail.com).  }
\thanks{G. Ricci is with Universit\`a del Salento, Via Monteroni snc, 73100 Lecce, Italy (e-mail: giuseppe.ricci@unisalento.it).  }
}

\maketitle
\begin{abstract}
In this work, we develop and compare two innovative strategies for parameter estimation and radar detection of
multiple point-like targets. The first strategy, which appears here for the first time, jointly exploits the
maximum likelihood approach and Bayesian learning to estimate targets' parameters including their positions
in terms of range bins. The second strategy relies on the intuition that for high signal-to-interference-plus-noise ratio 
values, the energy of data containing target components projected onto the nominal steering direction 
should be higher than the energy of data affected by interference only.
The adaptivity with respect to the interference covariance
matrix is also considered exploiting a training data set collected in the
proximity of the window under test. Finally, another important
innovation aspect concerns the 
adaptive estimation of the unknown
number of targets by means of the model order selection rules.
\end{abstract}


\IEEEpeerreviewmaketitle
\section{Introduction}
\IEEEPARstart{I}{n}
the recent years, advances in technology have paved the way for the design of radar processing units where 
high-performance sophisticated algorithms are executed to comply with the radar tasks and the system requirements.
Consider, for example, the space-time adaptive detection algorithms that exploit large volumes of data from sensor
arrays  and/or pulse trains to take advantage of temporal and spatial integration/diversity
\cite{Klemm,Richards} at the price of an increased computational load \cite{kelly1986adaptive,robey1992cfar,928688,RicciRao,
Yuri01,kraut1999cfar,8809199}. Another important example concerns high resolution radars,
which can resolve a target into a number of different scattering centers depending on the radar bandwidth and
the range extent of the target \cite{ScheerMelvin,928688,8679598}. In this case, conventional radar detection algorithms
process one range bin at a time despite the fact that contiguous cells can contain target energy.
As a consequence, they do
not collect as much energy as possible to increase the signal-to-interference-plus-noise ratio (SINR).
To overcome this drawback, architectures capable of detecting distributed targets by exploiting a certain
number of contiguous range bins have been developed \cite{Gerlach,DD,928688}.

The energy issues described for range-spread targets also hold for multiple point-like targets and
detection algorithms, which can take advantage of the total energy associated with all the point-like targets, are
highly desirable. However, the problem of jointly detecting multiple point-like targets is very difficult
since target positions along with target number (which deserves special attention) are unknown parameters that must be estimated.
Existing examples in the open literature share the assumption that the number of targets (or at least
an upper bound on it) is known and are based upon the maximum likelihood (ML) approach \cite{4014365,1605248}.

In this work, we devise two architectures to detect multiple point-like targets without assuming that the number of
targets is known. The first approach jointly exploits ML and Bayesian learning
\cite{887041} to estimate targets' parameters including their positions in terms of range bins. 
The second approach relies on the intuition that for high SINR values, the energy of data containing 
target components projected onto the nominal steering direction
should be higher than the energy of data affected by interference only (it can be shown that it is a special case
of \cite{4014365}). Thus, it is possible to discard range bins with low energy according to a given criterion.
In both cases, ad hoc modifications of the generalized likelihood ratio test (GLRT) are used to design adaptive architectures based upon
previous estimates.  Remarkably, the number of targets is
adaptively estimated by means of model order selection (MOS) rules \cite{Stoica1}, as the Bayesian information
criterion (BIC), the Akaike information criterion (AIC), and the generalized information criterion (GIC).

Finally, the performance analysis, conducted on simulated data and in comparison with classical detectors
for extended and/or multiple targets, is aimed at showing the effectiveness of the proposed architectures
in terms of both estimation and detection performance.

The remainder of this paper is organized as follows. In Section \ref{section2}, we formulate the detection problem 
at hand and define preliminary quantities which are used in Section \ref{secdet} to derive two adaptive procedures. 
Finally, in Section \ref{secNA} we assess the performance of the proposed methods in comparison with
their classical counterparts, whereas in Section \ref{secconclusion} we draw some concluding remarks\footnote{
\emph{Notation:}
In the sequel, vectors and matrices are denoted by boldface lower-case and upper-case letters, respectively. 
For a generic vector $\bx$, the symbol $\|\bx\|$ indicates its Euclidean
norm. The symbols $\det(\cdot)$, $(\cdot)^T$, and $(\cdot)^\dag$ denote the determinant, transpose, and conjugate transpose, respectively. 
As to numerical sets, $\C$ is the set of complex numbers, $\C^{N\times M}$ is the Euclidean 
space of $(N\times M)$-dimensional complex matrices (or vectors if $M=1$).
Let $x\in\C$, then $|x|$ denotes the modulus of $x$, whereas for each set $A$, $|A|$ stands for the cardinality of $A$.
The set difference is denoted by $\setminus$ while ${N}\choose{k}$ indicates the binomial coefficient.
Symbols $\bI$ and $\bzero$ represent the identity matrix and the null vector or matrix, respectively, of proper dimensions. 
Let $\bx$ a random vector, we denote by $f(\bx;\btheta)$ the probability density function (PDF) of $\bx$ with parameter vector $\btheta$.
If $C$ is an event and $\bx$ a random vector, $P\{C\}$ and $P\{C | \bx\}$ are the probability  of $C$ and the probability  of $C$ given $\bx$.
Finally, we write $\bx\sim\cC\cN_N(\bm, \bM)$ if $\bx$ is an  $N$-dimensional complex normal random vector with mean $\bm \in\C^{N\times 1}$ 
and positive definite covariance matrix $\bM\in\C^{N\times N}$.}.

\section{Problem Formulation}
\label{section2}
\begin{figure}
    \centering
    \includegraphics[width=0.35\textwidth]{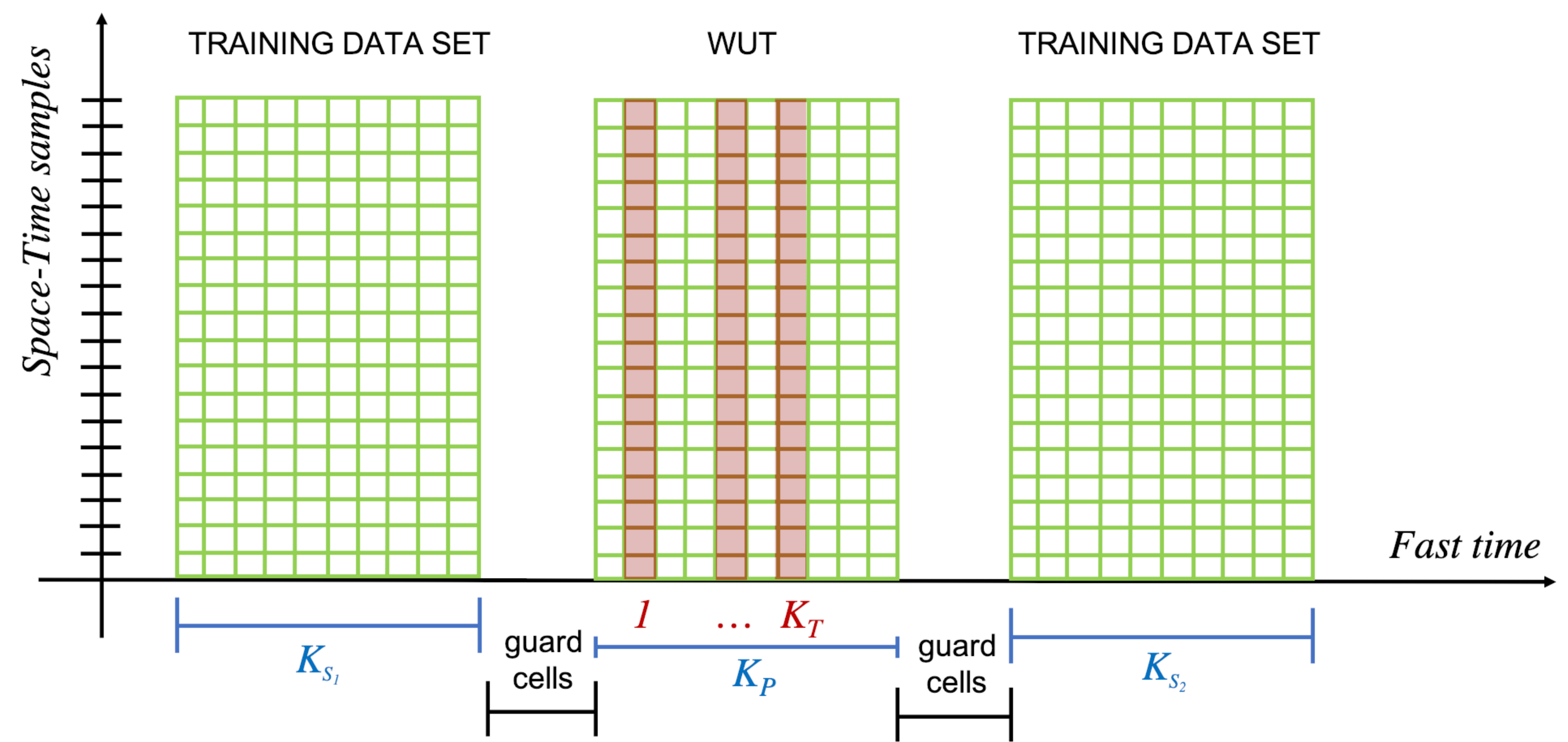}
    \caption{Window under test and training data set ($K_{S_1}+K_{S_2}=K_S$).}
    \label{fig:schema}
\end{figure}

Let us consider a radar system that transmits a burst of $N_p$ pulses by means of a linear array of $N_a$ antennas 
to sense the surrounding environment. 
The received echoes from the environment are suitably conditioned and organized to form $N$-dimensional vectors with $N=N_a N_p$
representing the range bins \cite{Richards,BOR-Morgan}. Thus, for each range bin belonging to the window under test (WUT), the corresponding
$N$-dimensional vector is the result of the superposition of an interference component (representative of thermal noise, clutter, etc.) 
and possible useful signal components. Besides, under the hypothesis that the environment is stationary over range and time
(the so-called homogeneous environment), a further set of training samples (secondary data) is collected by the system in proximity 
of the WUT and used to achieve adaptivity with respect to the interference covariance matrix (ICM) 
\cite{Richards,kelly1986adaptive,robey1992cfar,BOR-Morgan} 
(data sets to be processed are depicted in Figure \ref{fig:schema}).

Thus, denoting by
$\bZ=[\bz_1 \cdots \bz_{K_p}]\in\C^{N\times K_P}$ a matrix whose columns are the vectors belonging
to the WUT and by $\bR=[\bor_1 \cdots\bor_{K_S}]\in\C^{N\times K_S}$ the secondary data matrix,
we are interested in solving the following hypothesis test
\be
\left\{
\begin{array}{l}
H_1:
\begin{cases}
\begin{cases}
\bz_h\sim\cC\cN_N(\alpha_h\bv(\theta_T,\nu),\bM), & h\in\Omega_T,
\\
\bz_h\sim\cC\cN_N(\bzero,\bM), & h\in\Omega_P \setminus\Omega_T,
\end{cases}
\\
\bor_k\sim\cC\cN_N(\bzero,\bM), \quad\quad\quad k\in\Omega_S,
\end{cases}
\\
H_0:
\begin{cases}
\bz_h\sim\cC\cN_N(\bzero,\bM), & h\in\Omega_P,
\\
\bor_k\sim\cC\cN_N(\bzero,\bM), & k\in\Omega_S,
\end{cases}
\end{array}
\right.
\label{eqn:hypothesisTest}
\ee
where $\alpha_h\in\C$ is a complex factor accounting for the received energy backscattered by a coherent target;
$\bv(\theta_T,\nu)\in\C^{N\times 1}$ is the space-time steering vector with 
$\nu$ the normalized target Doppler frequency and $\theta_T$ the target angle of arrival measured with respect to
the array normal\footnote{In the remainder of the paper for simplicity, we omit the dependence of $\bv$ on $\nu$ and $\theta_T$.};
$\bz_1,\ldots,\bz_{K_P},\bor_1,\ldots,\bor_{K_S}$ are statistically independent random vectors;
$\Omega_P=\{ 1,\ldots,K_P \}$, $\Omega_S=\{ 1,\ldots,K_S \}$, and $\Omega_T\subseteq \Omega_P$ with $|\Omega_T|=K_T$.

It is important to underline here that the above problem describes an operating situation where the radar system jointly processes a set of contiguous
range bins which might contain $K_T\leq K_P$ point-like targets.
The main related problem is that $\Omega_T$ is unknown and must be somehow estimated from data.

Finally, for future developments, let us write the PDF of $\bz_h$ under $H_i$, $i=0,1$, as
\[
f(\bz_h;I_{\Omega_T}(h)i\alpha_h,\bM)=
\frac{e^{ -\|\bM^{-1/2}\left(\bz_h-I_{\Omega_T}(h)i\alpha_h\bv\right)\|^2 }}
{\pi^N \det(\bM)},
\]
where $I_{\Omega_T}(h)$ is the indicator function of $\Omega_T$. It follows that the joint PDF of $\bZ$ can be written as
\[
f_1(\bZ;\bM,K_T,\cA,\Omega_T)\!=\!\!\!\prod_{h\in\Omega_T} \!\!\! f(\bz_h;\alpha_h,\bM)
\!\!\! \!\!\! \prod_{h\in\Omega_P \setminus \Omega_T} \!\!\!\!\!\! f(\bz_h;0,\bM)
\] 
under $H_1$, where $\cA=\{ \alpha_h: \ h\in\Omega_T \}$,
and\footnote{Note that $\cA$ depends on $\Omega_T$, which is function of the actual number of targets. However,
we omit these dependences in order to simplify the notation.} as
$f_0(\bZ;\bM)=\prod_{h=1}^{K_P}f(\bz_h;0,\bM)$ under $H_0$.

\section{Adaptive Detector Design}
\label{secdet}
In this section, we drive the reader towards the design
of fully-adaptive detection architectures by gradually adding an adaptivity layer at each step.
As a matter of fact, in the first two subsections, both estimation
procedures  along with the related decision schemes 
are derived assuming that $K_T$ and $\bM$ (namely the ICM) are known (the remaining parameters
are clearly estimated from data). In the last subsection, we make the previously developed architectures
adaptive with respect to the ICM and $K_T$. To this end, the ICM is replaced by the sample covariance matrix,
whereas the adaptivity with respect to $K_T$ is achieved by resorting to the MOS rules, which allow us to estimate $K_T$.
Finally, the adopted detector design criteria rely on ad hoc modifications of the GLRT where the unknown
parameters are replaced by the estimates returned by the proposed procedures.

Precisely, the GLRT for known $\bM$ and $K_T$ (and based upon the range bins of the WUT) has the following expression
\be
\dmax_{\Omega_T}\dmax_{\cA}
\Lambda_1(\bZ;\bM,K_T,\cA,\Omega_T)\test\eta,
\label{eqn:GLRT}
\ee
where $\Lambda_1(\bZ;\bM,K_T,\cA,\Omega_T)={f_1(\bZ;\bM,\cA,K_T,\Omega_T)}$ $/ {f_0(\bZ;\bM)}$
and $\eta$ is the threshold\footnote{Hereafter, we denote by $\eta$ the generic detection threshold.}
to be set according to the desired probability of false alarm ($P_{fa}$).
Before moving to the heart of the derivations, we briefly outline the reasoning followed in this subsection.
Specifically, let us notice that solving problem \eqref{eqn:GLRT}
represents a difficult task (at least to the best of authors' knowledge)
due to the maximization with respect to $\Omega_T$.
Indeed, an exhaustive search 
would deal with ${K_P}\choose{K_T}$ subsets of $\Omega_P$, each with cardinality $K_T$.
Now, when $K_P/K_T$ becomes large, this approach would be prohibitive from a computational point
of view. For this reason, we conceive two alternative and ``smart'' strategies.


\subsection{Joint Bayesian and ML (BML) estimation}
\label{BML}

In the first proposed solution, we follow an alternative route looking at problem \eqref{eqn:GLRT} from a different perspective.
The strategy consists in defining hidden random variables (RVs) which represent the classes to which the range bins belong with 
a given probability. Specifically, we treat the estimation of $\Omega_T$ like a clustering problem, where
the class labeled as ``$0$'' corresponds to the absence of target while 
the bins containing prospective targets are associated with the classes
defined by indices greater than $0$. As a consequence, it is possible to
isolate the range bins containing interference only from the others. In this context, we assumes that each range bin belongs to
one of two classes, including labels ``$0$'' and ``$1$'', which refer to the absence and the presence of the target, respectively.
Specifically, let $b_h$, $h\in\Omega_P$, be independent RVs, with alphabet $\cB = \{ 0,1\}$ and the related probability mass function
$P\{b_h=k\}=\pi_k^{\textrm{BML}}$, $h\in\Omega_P$, $k\in\cB$.
Note that $b_h=1$, $h\in\Omega_P$,  if and only if the $h$-th range bin contains a target echo with the
consequence that, if $K_T$ targets are distributed over $K_P$ range bins, the resulting class priors are $\pi_1^{\textrm{BML}} = \frac{K_T}{K_P}$
and $\pi_0^{\textrm{BML}} = \frac{K_P-K_T}{K_P}$.
The PDF of $\bz_h$, $h\in\Omega_P$, under $H_1$, can be  written as a linear combination of two ``sub-PDF'', namely
$f(\bz_h; \bpi^{\textrm{BML}}, \alpha_h,\bM)= \pi_0^{\textrm{BML}}f(\bz_h;0,\bM) + \pi_1^{\textrm{BML}}f(\bz_h;\alpha_h,\bM)$,
where $\bpi^{\textrm{BML}}=[\pi_0^{\textrm{BML}},\pi_1^{\textrm{BML}}]^T$.
Now, we jointly exploit the Bayesian framework and the ML approach to come up with
an estimate of $r_{hk}^{\textrm{BML}}=P\{b_h=k|\bz_h\}$ for each $h\in\Omega_P$. Thus,
resorting to the Bayes rule, we can write
\be
r_{hk}^{\textrm{BML}} = {\pi_k^{\textrm{BML}} f(\bz_h; k \alpha_h,\bM)}
/{\ds \sum_{i=0}^{1}\pi_i^{\textrm{BML}} f(\bz_h;i\alpha_h,\bM)}.
\label{eqn:E_stepm}
\ee
As for the generic $\alpha_h$, we estimate it by solving a classical ML problem \cite{kelly1986adaptive} to obtain
$\widehat{\alpha}_h^{\textrm{ML}}={\bv^\dag \bM^{-1} \bz_{h}}/{\bv^\dag\bM^{-1}\bv}$.
Replacing this estimate in \eqref{eqn:E_stepm}, we obtain
\be \label{eqn:E_stepmrhk}
\widehat{r}_{hk}^{\textrm{BML}} = \frac{ \pi_k^{\textrm{BML}}  f(\bz_h;k\widehat{\alpha}_h^{\textrm{ML}},\bM)}{\ds \sum_{i=0}^{1}
\pi_i^{\textrm{BML}} f(\bz_h;i\widehat{\alpha}_h^{\textrm{ML}},\bM)}, \quad h\in\Omega_P, \quad k\in\cB.
\ee
Finally, $\widehat{r}_{hk}^{\textrm{BML}}$ can be suitably exploited
to obtain an estimate of $\Omega_T$, $\widehat{\Omega}_T^{\textrm{BML}}$ say, by discarding the
range bins indexed by $h\in\Omega_P$ such that
$\arg\dmax_{k} \widehat{r}_{hk}^{\textrm{BML}} = 0$
and, if the remaining set contains more than $K_T$ elements, by selecting the remaining $K_T$ range bins with the 
highest $\widehat{r}_{h1}^{\textrm{BML}}$. When the number of the remaining bins is less than $K_T$, we select
all of them.

Now, given the estimates of $\alpha_h$, $h\in\Omega_P$, and $\Omega_T$, we can build up the following decision rules
\be
\Lambda_1^{\textrm{BML}}(\bZ;\bM,K_T)=
\frac{\ds\prod_{h\in\widehat{\Omega}^{\textrm{BML}}_T} f(\bz_h; \widehat{\alpha}^{\textrm{ML}}_h,\bM)}
{\ds\prod_{h\in\widehat{\Omega}^{\textrm{BML}}_T} f(\bz_h;0,\bM)}\test\eta
\label{eqn:GMM_GLRT_03}
\ee
that is statistically equivalent to the generalized adaptive matched filter (GAMF)
computed over $\widehat{\Omega}_T^{\textrm{BML}}$ \cite{928688}.

\subsection{Estimation based upon energy}
\label{energybased}

From an intuitive point of view, when $\bz_h$ contains high-SINR target echoes, the energy of
the component along $\bv$ or $\bM^{-1/2}\bv$ should be higher than the component in the case where the target is not present. 
Otherwise stated, let $\bz_{h_1}=\alpha_{h_1}\bv+\bn_{h_1}$ and $\bz_{h_2}=\bn_{h_2}$, 
 with $\bn_{h_i}\sim\cC\cN_N(\bzero,\bM)$, $i=1,2$, then we expect that
$|\bv^\dag\bM^{-1}\bz_{h_1}|^2 > |\bv^\dag\bM^{-1}\bz_{h_2}|^2$.
The above insight suggests a strategy for the selection of the elements belonging to $\Omega_T$. Specifically,
let us order the range bins according to the energy amount of the component along $\bM^{-1/2}\bv$, namely
$|\bv^\dag\bM^{-1}\bz_{h_1}|^2 > |\bv^\dag\bM^{-1}\bz_{h_2}|^2 > \ldots > |\bv^\dag\bM^{-1}\bz_{h_{K_P}}|^2$
and, then, select the first $K_T$ range bins $\bz_{h_1},\ldots,\bz_{h_{K_T}}$ to form  
$\widehat{\Omega}_T^{\textrm{WEN}}$  (where WEN stands for Whitened ENergy).
Finally, such estimates can be exploited to
come up with the following decision
rule\footnote{Note that $\Lambda_2^{\textrm{WEN}}(\bZ;\bM,K_T)$ is the GAMF computed over $\widehat{\Omega}_T^{\textrm{WEN}}$.}
\begin{align}
\Lambda_2^{\textrm{WEN}}(\bZ;\bM,K_T) &= \sum_{k\in\widehat{\Omega}_T^{\textrm{WEN}}} \frac{|\bv^\dag\bM^{-1}\bz_{k}|^2}{\bv^\dag\bM^{-1}\bv}
\test\eta.
\label{eqn:GAMF_Wenergy}
\end{align}


\subsection{Adaptivity with respect $\bM$ and $K_T$}
Assuming that the ICM is known does not have a practical value. As a matter of fact, the 
a priori information about the ICM is often limited to its specific structure. Such information might come
from possible symmetries induced by system geometry or clutter properties \cite{CP00,DeMaioSymmetric,fogliaPHE_SS,HaoSP_HE}.
For this reason, in order to make architectures \eqref{eqn:GMM_GLRT_03} and \eqref{eqn:GAMF_Wenergy} adaptive with
respect to $\bM$, we replace the latter with the ubiquitous sample covariance matrix based upon 
secondary data set \cite{muirhead2009aspects,robey1992cfar,BOR-Morgan}, namely $\widehat{\bM}=\frac{1}{K_S} \bR\bR^\dag$.

The final step towards adaptivity consists in estimating $K_T$ from the observables. In fact, there exist situations
where the a priori information about the actual number of targets
is not available or is not exact. 
For this reason, we resort to the MOS rules whose general expression is
\be
\widehat{K}_T=\underset{k=1,\ldots,K_P}{\arg\min} -2\log f_1(\bZ;\widehat{\bM},k,\widehat{\cA}(k),\widehat{\Omega}_T(k))+p(k).
\label{eqn:MOS}
\ee
In the above equation, $\widehat{\cA}(k)$ and $\widehat{\Omega}_T(k)$ are the
estimates\footnote{Actually MOS rules exploit the compressed likelihood function, 
where the parameters are replaced
by the respective ML estimates. In our development, we replace the parameters with alternative estimates
when the former are not available.} 
of $\cA$ and $\Omega_T$, respectively, assuming that $K_T=k$;
$p(k)=3 k \nu$ is the penalty term \cite{Stoica1}, 
where the factor $3$ represents the number of unknowns for each target, i.e., 
the complex-valued amplitude and its position, while $\nu=2$ for AIC, 
$\nu=(1+\rho)$, $\rho\geq 1$, for GIC, and $\nu=\log K_P$ for BIC \cite{Stoica1}.

Finally, we obtain the adaptive detectors replacing the unknown matrix $\bM$ and $K_T$ in \eqref{eqn:GMM_GLRT_03}
and \eqref{eqn:GAMF_Wenergy} with the sample covariance matrix and \eqref{eqn:MOS}, respectively.


\section{Simulation Results}
\label{secNA}

In this section, we analyze the performance of the newly proposed unsupervised learning approaches 
resorting to standard Monte Carlo (MC) counting techniques. More precisely,
the detection performance and the thresholds are estimated over $10^3$ and $100/P_{fa}$ independent trials, respectively.
In all the illustrative examples, we set $N=N_a=16$, $K_T=5$, $K_P=30$, $K_S=48$, and $P_{fa} = 10^{-3}$.
The ICM is given by $\bM=\bI+\textrm{CNR}\ \bM_c$, with a clutter to noise ratio (CNR) of 40 dB.
The $(i,j)$th entry of the clutter component $\bM_c$ is given by $\bM_c(i,j)=\rho_c^{|i-j|}$ with $\rho_c=0.9$.
The SINR is defined as $\textrm{SINR}=P_{\alpha}^{av}\bv(0,0)^{\mathrm{\dag}}\bM^{-1}\bv(0,0)$, where $P_{\alpha}^{av}$ is the average power of the point-like targets.

\begin{figure}[htp!]
    \centering
    \includegraphics[width=.44\textwidth]{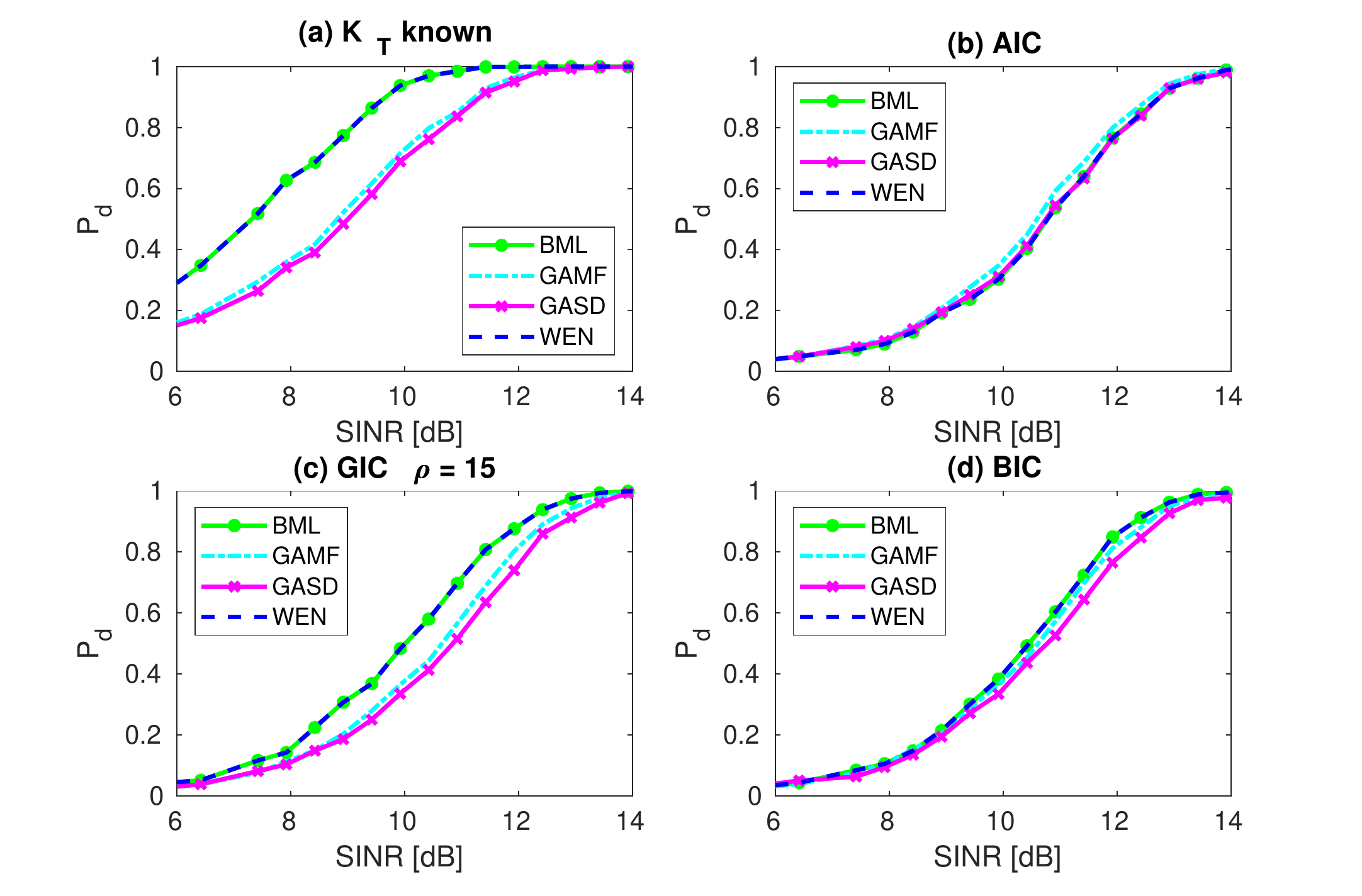}
    \caption{$P_d$ versus SINR for \eqref{eqn:GMM_GLRT_03}, \eqref{eqn:GAMF_Wenergy}, the GAMF, and the GASD assuming $K_P=30$ and
     $K_S=48$.}
    \label{fig:Pd_fixed}
\end{figure}

In Figure \ref{fig:Pd_fixed}(a), we plot the probability of detection ($P_d$) versus SINR
for \eqref{eqn:GMM_GLRT_03} and \eqref{eqn:GAMF_Wenergy} with known $K_T$.
For comparison purposes,
we also report the $P_d$ curves of the so-called GAMF and GASD introduced in \cite{928688}. It is possible to observe
that both the proposed procedures are capable to achieve a gain close to $1.6$ dB
at $P_d=0.9$ with respect to the competitors. The remaining subfigures refer to the case where $K_T$ is unknown.
In this case, due to the estimation of the latter, the gain over the GAMF and GASD becomes lower than or equal to about $1$ dB
for all the considered architectures. Specifically, the maximum gain is achieved by the GIC-based architectures, whereas
the gain associated with the BIC-based detectors decrease to about $0.5$ dB; the AIC-based decision schemes share the same performance as the GASD.
 Finally, notice that the similar detection performance of \eqref{eqn:GMM_GLRT_03} and \eqref{eqn:GAMF_Wenergy} 
can be explained by the fact that, as shown below, $\widehat{\Omega}_T^{\textrm{WEN}}$ and $\widehat{\Omega}_T^{\textrm{BML}}$ contain almost
the same indices.

The estimation quality of both MOS-based procedures is assessed 
in Figures \ref{fig:rms_mos} and \ref{fig:histo_mos}.
In Figure \ref{fig:rms_mos}, we consider the following
figures of merit as functions of the SINR:
the root mean square (RMS) number of missed targets evaluated by verifying that the estimated 
vector $\widehat{\balpha}= [\widehat{\alpha}_1 \ \cdots \ \widehat{\alpha}_{K_p} ]^T$ has zero elements in the actual target
positions;
the RMS number of ghosts (false positive targets) defined as the 
non-zero components of $\widehat{\balpha}$ in positions different from that of the targets;
the Hausdorff metric \cite{4567674} between $\balpha=[ \alpha_1 \ \cdots \ \alpha_{K_p} ]^T$
and $\widehat{\balpha}$. In Figure \ref{fig:histo_mos}, 
we show the histograms that represent the probability of selecting a specific
number of targets over $1000$ MC trials (assuming two different SINR values).
Inspection of these figures highlights that both proposed estimation procedures share the same performance
since the values of the considered metrics are overlapped. Moreover, AIC- and BIC-based
procedures experience a floor for the number ghosts and Hausdorff metric leading to poor estimation performance
as corroborated by the histograms. On the contrary, GIC-based procedures are capable of returning the correct
number of targets with high probability ($>0.75$).
For the above reasons, the recommended architectures for the 
detection problem at hand are those relying on GIC.

\section{Conclusion}
\label{secconclusion}
The problem of parameter estimation and adaptive detection for multiple point-like targets has been addressed
using two different approaches. In the first case, we jointly resorted to ML estimation and Bayesian
learning. 
The second approach is based on energy considerations. Adaptivity with respect to the ICM and the number of
targets is achieved through the exploitation of a training data set and of the MOS rules, respectively.
The illustrative examples have shown the superiority of the GIC-based approaches with respect 
to the other MOS-based solutions and to the competitors.
\begin{figure}[htp!]
    \centering
    \includegraphics[width=.41\textwidth]{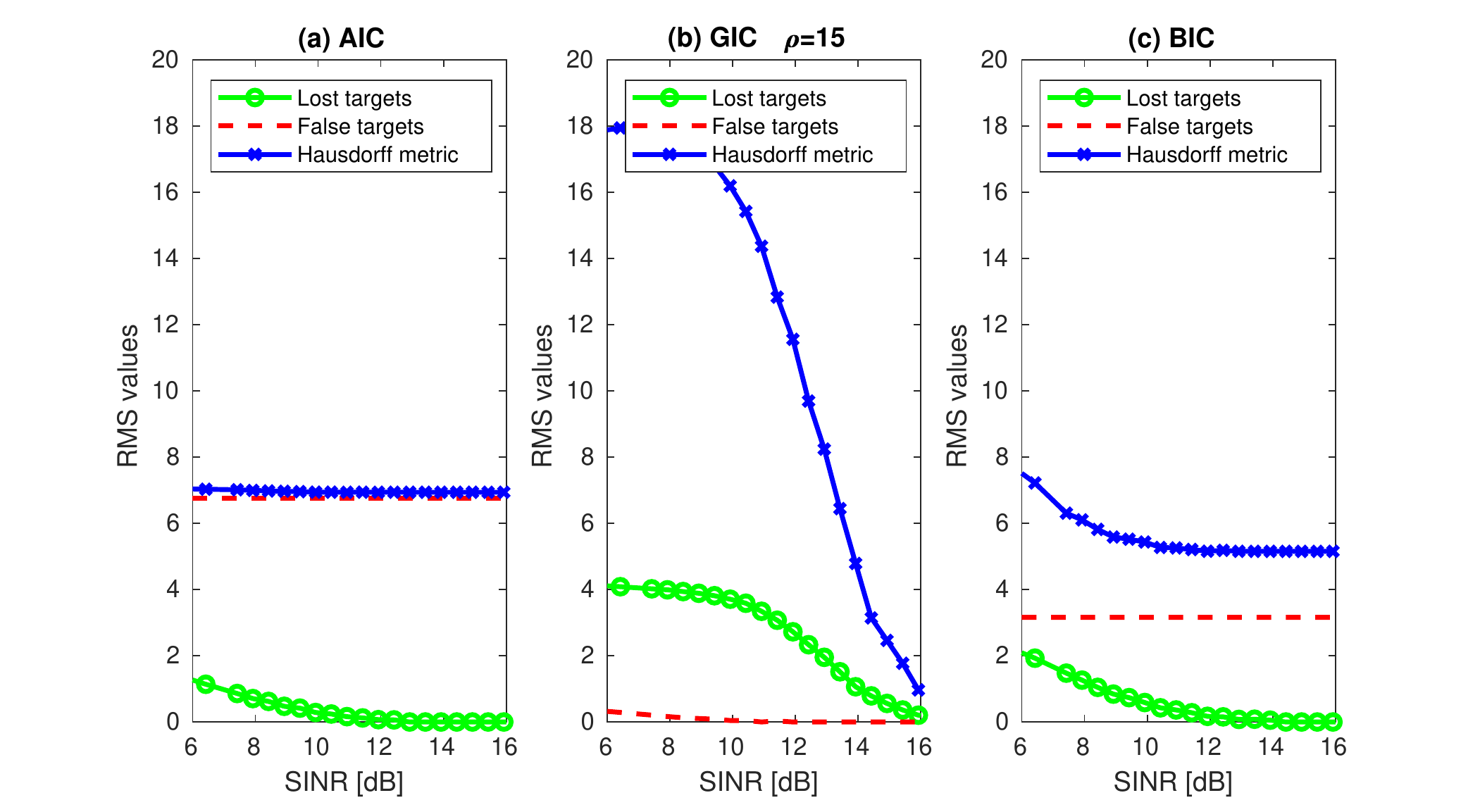}
    \caption{RMS values versus SINR for both procedures assuming $K_P=30$ and $K_S=48$.}
    \label{fig:rms_mos}
\end{figure}
\begin{figure}[htp!]
    \centering
    \includegraphics[width=.32\textwidth]{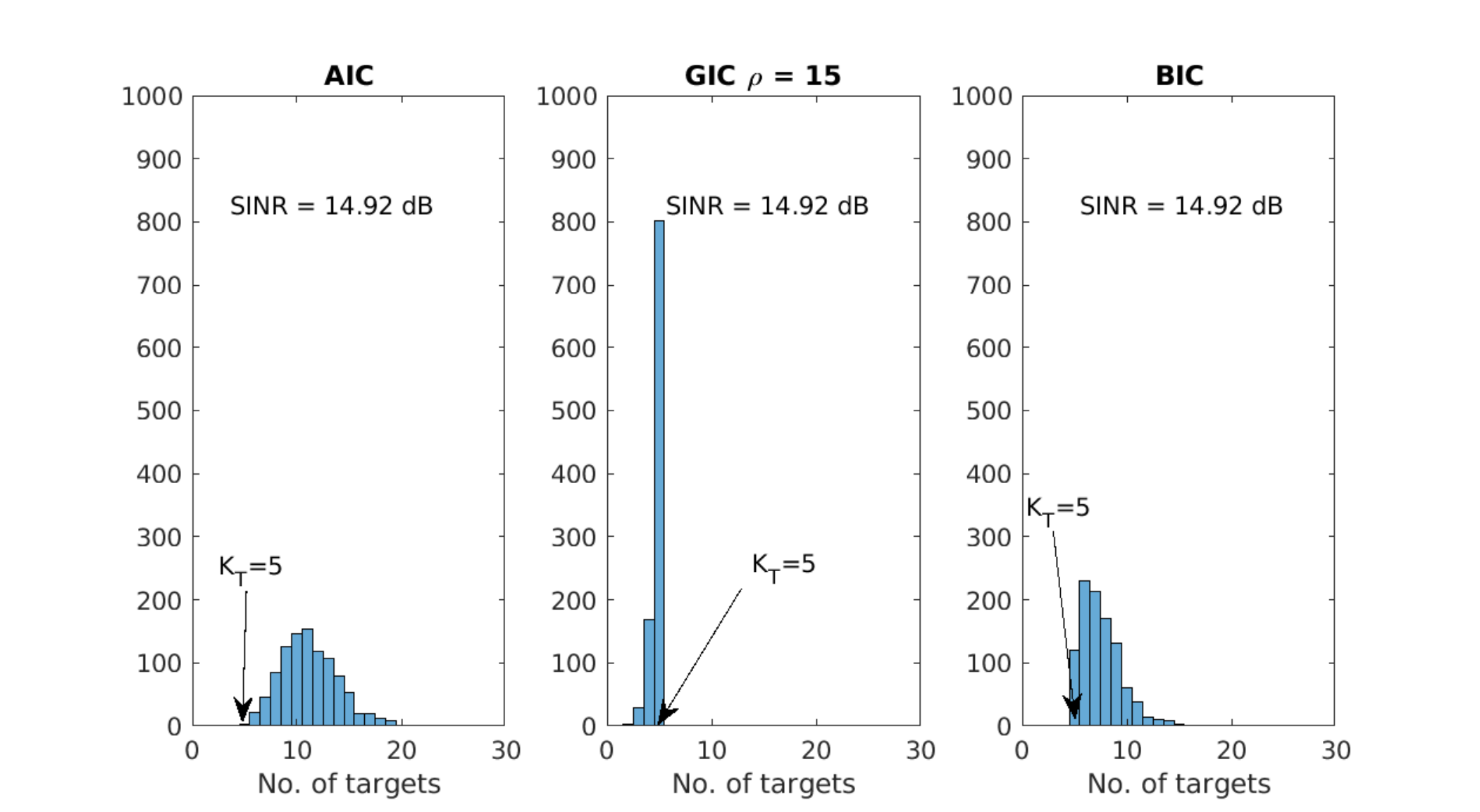}
    \includegraphics[width=.32\textwidth]{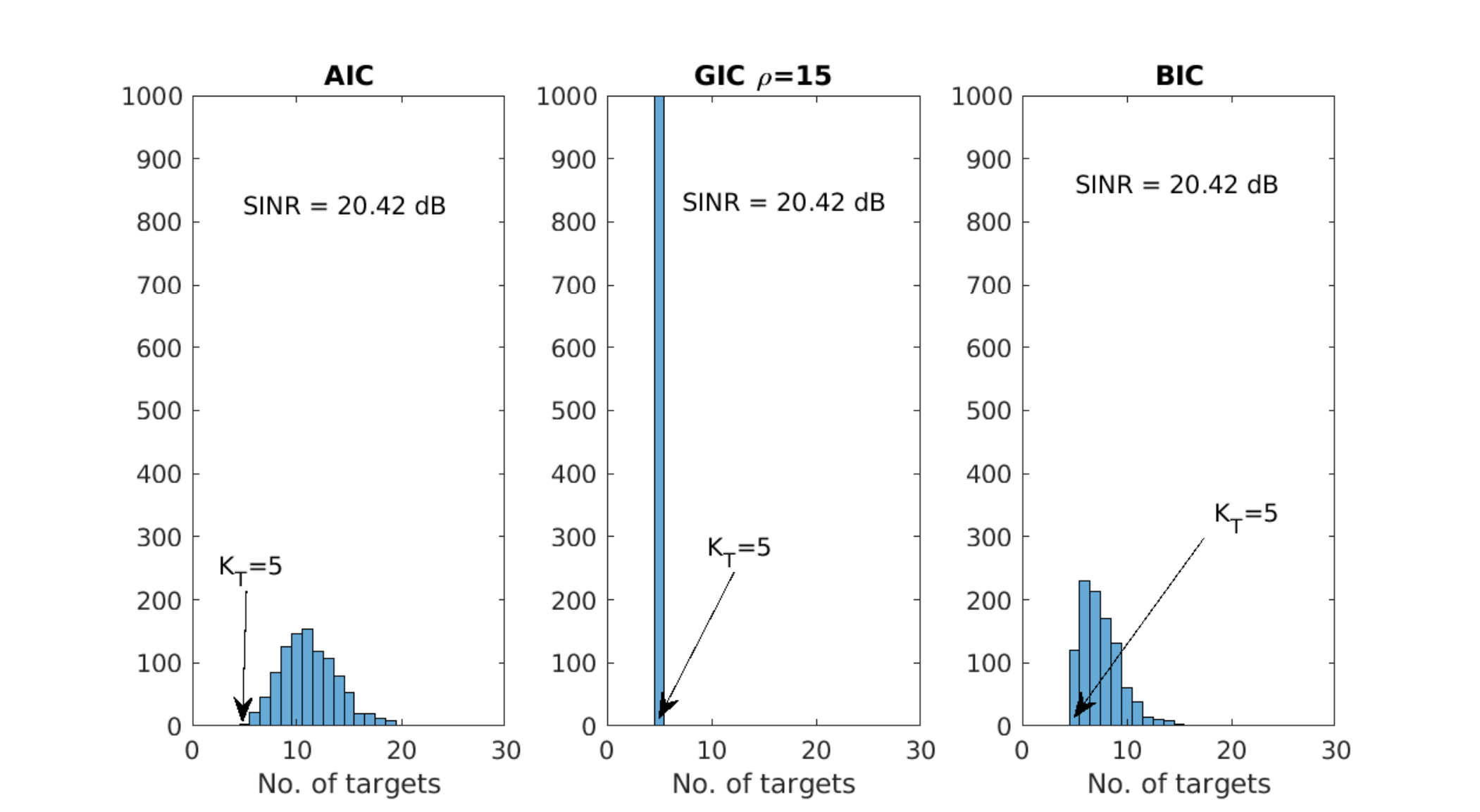}
    \caption{Histograms of the occurrence of $\widehat{K}_T$ values assuming $K_P=30$, $K_T=5$, and $K_S=48$.}
    \label{fig:histo_mos}
\end{figure}
%
%
\balance
\bibliographystyle{IEEEtran}
\bibliography{group_bib}
\end{document}